# Deep Mediterranean turbulence motions under stratified-water conditions

by Hans van Haren

Royal Netherlands Institute for Sea Research (NIOZ), P.O. Box 59, 1790 AB Den Burg, the Netherlands.
e-mail: hans.van.haren@nioz.nl

**Abstract.** Vertically stable in density, stratified-water conditions 'SW' exist in the deep Mediterranean Sea that are characterized by temperature differences of 0.0002-0.01°C over 125 m above a flat seafloor. These result in a mean buoyancy frequency of N = (1.5-2)f, where f denotes the inertial frequency. Although the stability values are one order of magnitude smaller than found in the ocean, they govern a dynamical deep sea as demonstrated using observations from a 3D mooring-array equipped with nearly 3000 high-resolution temperature sensors. SW-conditions can last up to a fortnight, before waters become near-homogeneous, and occur about 40% of the time, slightly more often in winter than in summer. Under SW, up to 60 m above seafloor is dominated by convection turbulence that is partially driven by geothermal heating 'GH' suppressed by stratification above. The upper-half of the array shows dominant shear turbulence driven by two sources. Interfacial internal waves generate weakly-nonlinear, resonant parametric instabilities that, upon breaking, provide mean turbulence dissipation rates of about one-third of that via general GH. It is about equal to open-ocean values away from boundaries and may represent the dominant source of turbulence there. Like GH, the observed turbulence is local up- and down-going. Tenfold larger mean dissipation rates are observed when slanted convection drives turbulent overturns >10 m and unstable clouds are advected with the mean flow. It confirms theoretical marginal stability analyses, previous vertical waterflow observations, and suggests a relationship between turbulence and sub-mesoscale eddies across the internal wave band. Movies support the findings.

# 1 Introduction

In the over 2000-m deep Western Mediterranean Sea milli- to centi-degree variations in temperature characterize all dynamical processes. This deep sea may be void of sunlight and its waterflow may be relatively slow, it is not stagnant but requires precise instrumentation for studying. Above the generally flat seafloor in the vicinity of its northern continental shelf the dynamical processes include a large-scale boundary flow (e.g., Crepon et al., 1982), with meanders and eddies varying at 1-10 day and 1-10 km sub-mesoscales, as well as at 10-30 day and 10-100 km mesoscales. The eddies result from instabilities of the boundary flow and associated fronts, horizontal density variations. These variations are strongest near the surface, but can be found all the way to the seafloor in weaker form. Shorter-period variations involve internal waves, notably at near-inertial scales, as transients following passages of atmospheric disturbances induced, e.g., by varying winds from the nearby mountain ranges like the Alps. Near-inertial motions may be trapped by the (sub-)mesoscale eddies, and set-up shorter scale internal waves that eventually dissipate their energy through irreversible turbulence when they break. Although the lack of substantial tides reduces the amount of internal wave energy by about 60% (Wunsch and Ferrari, 2004), the existing motions make the Mediterranean Sea a sample for ocean-dynamics processes (Garrett, 1994).

Detailed observations of these dynamics processes and the quantification of the turbulence generated in the deep sea are scarce. Especially also, because at large depths all dynamics occur under weakly stratified conditions. The local mean buoyancy frequency in the deep Western Mediterranean has typical values of $N = O(f)$, f denoting the inertial frequency or vertical Coriolis parameter (van Haren and Millot, 2003; see also the local standard-oceanographic shipborne CTD-profile observations in Fig. 0a, b). Under such weakly stratified conditions internal waves extend well into the sub-inertial sub-mesoscale range as not only gravity but also Earth's rotational momentum play a role as restoring force (LeBlond and Mysak, 1978): inertio-gravity waves 'IGW'. While internal waves are difficult to distinguish from instrumental noise, some appear as wiggles O(10) m in the classic oceanographic profiles. Unstable wiggles may represent turbulent overturns although these are overwhelmed by noise in density anomaly (Fig. 0b).

Very weakly stratified conditions may extend several 100's of meters above the seafloor. In this range a large three-dimensional '3D' mooring-array has been deployed with nearly 3000 high-resolution temperature 'T-'sensors measuring nearly half a cubic hectometer for over one year. It was constructed to study 3D development of turbulence generation, observing various turbulence forms in short movies (van Haren et al., 2026), and improve deep-sea turbulence statistics, also under weakly stratified conditions. As the conditions push limits to the instrumentation, extensive post-processing is required to capture the physical signals.

In this paper, observations over a 17-day long period of deep-sea temperature development are presented under relatively stratified-water 'SW' conditions. It contrasts with observations under near-homogeneous conditions presented elsewhere (van Haren, 2026 submitted). Here, SW is defined as a daily-averaged (pressure-corrected) temperature difference larger than 0.0002°C between uppermost and lowest T-sensors at nominally $h$ = 125.5 and 1.5 m above the seafloor, respectively. For dynamics studies of the deep sea a relationship is required between temperature and density (Fig. 0a,b). The stable temperature (density) stratification results in a mean value of about $N$ = (1.5-2.5)$f$. The 17 days is about the longest consecutive period of SW in a yearlong record as demonstrated from a standard oceanographic single-line mooring (Fig. 0c).

The 3D large-ring mooring goals are to gain more statistics, improved physical insight and 3D development of 'small-scale', internal wave and turbulent, water motions under SW conditions in the deep sea. Anticipated dominant turbulence-dynamics processes are breaking internal waves, possibly steered by sub-mesoscale and near-inertial motions, yielding instabilities advected from above and sideways. As will be quantitatively demonstrated, averaged internal-wave turbulence varies by half-one order of magnitude in energy compared with that induced by geothermal heating 'GH' through the seafloor from below. The larger O(10-100) m internal-wave turbulent overturns are expected to be buoyancy driven, while the smaller O(1-10) m instabilities are expected to be driven via disintegration of a single harmonic internal wave into continuous spectra via parametric instability (Davis and Acrivos, 1967). The latter is a weakly nonlinear resonant interaction, while the former may result from a strong non-resonant interaction.

## 2 Several definitions

The traditional frequency ($\omega$) range for freely propagating internal waves is $f < \omega < N$, for $N \gg f$ (LeBlond and Mysak, 1978). In this range, motions near f are circularly polarized in the horizontal plane. $N^2$ is considered the 'large-scale mean' value of stable vertical density stratification, averaged over at least the inertial period and typically 100-m vertical scales that include all turbulence overturning scales. However, observed stratification is not uniformly layered and highly varies with time 't' and space 'x,y,z', being, for example, deformed by the passage of internal waves it supports.

Thus, $N_{min} < N(x,y,z,t) < N_m < N_{max}$, where minimum $N_{min}$ and maximum buoyancy frequency $N_{max}$ are defined for a given area and period in time. Hereby, somewhat arbitrary definitions are used for the vertical scale, which is limited by the separation distance of instrumentation (sampling) rather than by the physical 0.01-0.001 m turbulence dissipation scale in water. Practically in the present data, smallest and largest vertical scales are 2 and 124 m, respectively (Section 3). As $N_{max}$ and $N_{min}$ values rarely occur, in principle only once in a record, their effect on internal-wave band extension is rather small. A more practical extension is the mean $N_m$ of maximum (small-scale) buoyancy frequencies calculated per vertical profile every time step of sampling. Although SW-conditions are characterized by weak mean N, one order of magnitude larger $N_m$ (and hence $N_{max}$) are observed, also under these conditions.

Like mean N being non-constant in the ocean and deep sea, the local vorticity of horizontally rotating motions may vary around latitudinal($\varphi$)-dependent planetary value $f = 2\Omega\sin\varphi$. $\Omega$ denotes the Earth rotational frequency. Time- and space-varying vorticity induced by horizontal waterflow differences such as in eddies can generate relative vorticity of up to $|\zeta| = f/2$ near the surface in the Western Mediterranean (Testor and Gascard, 2006). This addition to planetary vorticity f automatically widens the 'effective' near-inertial band to $0.5f < f_{eff} < 1.5f$. Besides this modification of local near-inertial frequency (Perkins, 1976), anticyclonic eddies can trap downward propagation of near-inertial waves (Kunze, 1985). Such frequency modification may add to local physics of inertial-wave caustics due to latitudinal $\varphi(y)$ variation (LeBlond and Mysak, 1978), which however can maximally lead up to 15% change in f in the Mediterranean due to the limited size of the basin in meridional y-direction.

Probably the largest extension of the internal-wave band also under here studied SW-conditions is associated with the effect of traditionally neglected horizontal Coriolis parameter $f_h = 2\Omega\cos\varphi$. For $N = O(f)$, minimum IGW-bound $\omega_{min} \leq f$ and maximum IGW-bound $\omega_{max} \geq 2\Omega$ or $N$, whatever is largest, are functions of $N$, latitude $\varphi$ and direction of wave propagation (LeBlond and Mysak, 1978; Gerkema et al., 2008),

$$\omega_{max}, \omega_{min} = (A \pm (A^2-B^2)^{1/2})^{1/2}/\sqrt{2}, \qquad (1)$$

in which $A = N^2 + f^2 + f_s^2$, $B = 2fN$, and $f_s = f_h\sin\alpha$, $\alpha$ the angle to $\varphi$. For $f_s = 0$ or $N \gg 2\Omega$, the traditional bounds $[f, N]$ are retrieved from (1). One of the effects of $f_h$ is turbulent-convection in slantwise direction in the vertical, meridional $z, y$-plane, so that apparently stable stratification observed in strictly $z$-direction may actually reflect homogeneous or unstable conditions in the tilted plane of planetary vorticity, except at the north pole (e.g., Straneo et al., 2002; Sheremet, 2004).

For fixed local $f_{eff}$, the minimum absolute value of less than half-an-order of magnitude wide IGW-bandwidth $\omega_{max}$-$\omega_{min}$ is found under conditions of $N = 1f$, for propagation in meridional direction. The IGW thus widens under more and under less stratified conditions. For example, a one-order of magnitude IGW bandwidth is found approximately under $N = 8f \approx N_{max}$, in the deep Mediterranean.

## 3 Materials and Methods

The half-cubic-hectometer of deep Mediterranean seawater was measured every 2 s using 2925 self-contained high-resolution NIOZ4 T-sensors, which can also record tilt and compass. Temperature-only sensors were taped at 2-m intervals to 45 vertical lines 125-m tall (van Haren et al., 2021). In addition, two tilt-temperature sensors were attached near top and bottom of each line, which was tensioned to 1.3 kN by a single buoy above. Three buoys, equally distributed over the mooring-array, held a single-point Nortek AquaDopp current meter CM that recorded waterflow at a rate of once per 600 s. The lines were attached at 9.5-m horizontal intervals to a steel-cable grid that was tensioned inside a 70-m diameter steel-tube ring, which functioned as a 140-kN anchor. This 'large-ring mooring' was deployed at the <1° flat and 2458-m deep seafloor of 42° 49.50′N, 006° 11.78′E, 10 km south of the steep continental slope in the NW-Mediterranean Sea, in October 2020.

Probably due to a format error, the T-sensors switched off unintentionally after maximum 20-months of data-recording. Tilt-temperature sensors recorded only 5.5 months of data, which data are not considered here. As with previous NIOZ4 T-sensors (for details see van Haren, 2018), the T-sensors' individual clocks were synchronized to a single standard clock every 4 hours, so that all T-sensors were recording within 0.01-s precision. About 25 T-sensors failed mechanically. After calibration, some 20 extra T-sensors are not further considered due to general electronics (noise) problems. Data from these sensors are not considered in spectral analyses, and result from interpolation between neighbouring sensors in other analyses. During post-processing, specific limitation of the instrumentation was revealed concerning 'short-term bias'. That bias was removed via interpolation of a further 50-150 records depending on the type of analysis, elaborated below, and via vertical low-pass filtering 'lpf'.

Common, primary, 'long-term' electronic drift or bias of typically $0.001°C\ mo^{-1}$ after aging in NIOZ4 was corrected by referencing daily-averaged vertical profiles, which must be stable from turbulent-overturning perspective in a stratified environment, to a smooth polynomial without instabilities (van Haren et al., 2005; van Haren and Gostiaux, 2012). In addition, because vertical temperature (density) gradients are so small in the deep Mediterranean, reference was made to periods of typically one hour duration that were homogeneous with temperature variation smaller than instrumental noise level (van Haren, 2022). Such periods were found on days 350, 453, and 657 in the existing records. This secondary correction allowed for proper calculations of turbulence values using the Thorpe (1977) method of reordering unstable vertical density (temperature) profiles under weakly stratified conditions. For comparison, turbulence values were also calculated using the method of Ellison (1957).

With application for atmospheric observational data, Ellison (1957) separated time series of potential temperature $\theta(t, z)$, at a fixed vertical position in two,

$$\theta = <\theta> + \theta', \qquad (2)$$

where $<.>$ denotes the lpf time series and the prime its high-pass filtered equivalent. With multiple sensors deployed in the vertical for establishing a mean vertical temperature gradient, a root-mean-square 'Ellison-'scale can be defined as,

$$L_E = <\theta'^2>^{1/2}/(d<\theta>/dz). \qquad (3)$$

Like Thorpe (1977) displacement scales $L_T$, the $L_E$ may be compared with the Ozmidov (1965) scale $L_O$ of largest possible turbulent overturns in stratified waters, so that the turbulence dissipation rate reads,

$$\varepsilon = c^2 L_E^2 N^3, \tag{4}$$

in which the constant c needs to be established as a mean from a large distribution of values. If one takes an average value of c = 0.8 (Dillon, 1982) as follows from comparison of $L_T$ with $L_O$, $L_E \approx L_T$ was found under well-stratified deep-ocean conditions (Cimatoribus et al., 2014).

While the Thorpe (1977) method is most sensitive for displacements due to the largest vertical overturn scales, the Ellison (1957) method applied to moored T-sensor time series is most sensitive for the appropriate separation between internal waves and turbulent motions (Cimatoribus et al., 2014). This separation is not a straightforward task under weakly stratified conditions that occur in the deep Western Mediterranean. The method requires precise band-pass filtering of turbulent motions, excluding data governed by instrumental noise as well as internal waves from the turbulence records.

Under SW conditions, a tertiary correction involves filtering of data. Three phase-preserving double elliptic filters are applied (Parks and Burrus, 1987). One lpf is used per record in time with cut-off at 3000 cpd (cycles per day). Specific for the Ellison (1957) method, a high-pass filter is used with cut-off at twice the mean $N_m$ of maximum 2-m small-scale buoyancy frequencies that approximately amounts $N_m$ = 4.5-6 cpd. The third filter is used per line in the vertical with cut-off at about 0.2 cpm (cycles per metre). That lpf is designed for tertiary drift correction by removal of short-term bias, which manifests in records shorter than one day under weakly stratified conditions in water. This drift is likely associated with weak variable response to temperature changes. As will be demonstrated in Section 4, short-term drift of typically <0.0001°C acts within the time/frequency range of internal waves and turbulence, so that tertiary correction is necessary also before spectral analysis. Primary corrections are not necessary for spectral internal-wave and turbulence investigations, because instrumental drift manifests itself as slow variations with time, with typical scales of a month, well outside daily and shorter time scales.

Time-lpf data are also used to generate 3D movies, after forcing the data of lowest T-sensors to a common mean value because no drift correction is possible between different lines. Movies are created from a sequence of individual *.jpeg images using freeware video-utility VirtualDub (https://www.virtualdub.org/; last accessed 22 September 2025). They are accelerated, by a factor of up to 1800, with respect to real-time.

The large number of T-sensors in a small sea domain are also expected to improve statistics of at least part of turbulence motions. As temperature-variance spectra are commonly rather featureless for deep-sea observations, with small peaks and no gaps, the focus is on comparison of heavily-smoothed (averaged) spectra with turbulence models that generally describe a certain frequency band across which the variance $P(\omega) \propto \omega^p$ varies with slope p. This comparison is easiest performed when spectra are plotted on a log-log scale, in which p forms straight lines. Two main turbulence models are of interest, the Kolmogorov (1941) – Obukhov (1949) 'KO' model and the Bolgiano (1959) – Obukhov (1959) 'BO' model.

The KO-model describes the equilibrium inertial subrange of turbulence (Tennekes and Lumley, 1972) with a balance of turbulence kinetic energy 'KE' production and dissipation. It describes a forward cascade of energy, with turbulence dominantly produced by vertical current shear in which scalars are passively transported (Warhaft, 2000). As a result, the spectral slopes for KE and a scalar like temperature (variance) are identical in the inertial subrange, and amount p = -5/3.

The BO-model describes the buoyancy subrange of convection turbulence, which is actively driven by scalar density (or temperature). Its direction of energy cascade is still under debate and presumed partially forward and partially backward, as established for laboratory turbulence (Lohse and Xia, 2010). Its spectral slopes are p = -11/5 for KE and p = -7/5 for scalars like temperature.

For a nearly raw single-sensor record the spectral number of degrees of freedom 'dof' is so low, ~3 dof, that the KO- and BO-slopes are only significantly distinguishable over a frequency range of about three orders of magnitude. For near-random signals, improved statistics (~200 dof) for the average spectrum over an entire mooring line distinguishes the two spectral models over a range of about 1.2 orders of magnitude (e.g., Jenkins and Watts, 1968). Statistics over the 3D mooring-array theoretically allows distinction of the models over less than one-fifth order of magnitude.

The KO- and BO-turbulence models are also compared with various other spectral-slope models, including intermittency 'Im' of chaotic processes characterized by p = -1 (Schuster, 1984), and internal wave 'IW' that under well-stratified conditions is characterised by p = -2 for frequency range f << ω << N (Garrett and Munk, 1972). The p = -2 slope also describes finestructure contamination in records of finite length (Phillips, 1971; Reid, 1971).

As for dominant processes governing turbulence dissipation rate in the area, Ferron et al. (2017) calculated three times larger values due to GH than due to IGW-breaking. Their calculations were based on sparse shipborne microstructure profiling, extensively throughout the Western Mediterranean. For reference in periods with dominant GH, the local mean geophysical-determined heat flux amounts 0.11 W $m^{-2}$ (e.g., Pasquale et al., 1996). After conversion of this mean heat flux into buoyancy flux 'Bfl' in the overlying waters, mean GH-induced turbulence dissipation rate is calculated as,

$$\varepsilon_{GH} = Bfl/\Gamma_C = 1.2\times10^{-10} \text{ m}^2 \text{ s}^{-3}. \qquad (5)$$

In previous data (van Haren, 2025) the mixing coefficient was found to amount $\Gamma_C = 0.5$, which is typical for convection turbulence (Dalziel et al., 2008; Ng et al., 2016).

**4 Results**

Despite the applied de-trending, long-term instrumental drift of about 0.0005°C would not be well visible from the two temperature records in the SW data-overview of single-line time series (Fig. 1a). (Here and elsewhere, 'temperature' is used in short for pressure-corrected Conservative Temperature (IOC et al., 2010). While 4-7-day variations of about 0.004°C are seen, <0.5-day variations generally stay within 0.001°C. Between the two records obtained vertically 124-m apart, up to 0.006°C differences occur. Generally, the larger the vertical temperature difference, the warmer the waters, including at the lowest T-sensor. Comparison with time series of data from other quantities yields inconclusive results.

Waterflow speed (Fig. 1b) hardly shows quasi-inertial (0.73-day) variations. More dominantly, 2-3-day variations are observed that are difficult to relate with temperature variations. To some extent relative acoustic amplitude 'dI' (Fig. 1c) correlates with zero-lag with temperature variations. In general under SW-conditions, dI is 5-10 dB larger than under near-homogeneous conditions. Nearby-island wind variance (Fig. 1d) shows a broad peak, which may associate with the largest temperature difference after employing a 4-day advance. Daily (and vertically) averaged turbulence dissipation rate (Fig. 1e) shows hardly 2-4-day variations. All values except one exceed that of mean GH-value (5) of which the logarithm amounts -9.92. The overall mean turbulence dissipation rate for the 17-day period amounts $4\pm2\times10^{-10}$ m$^2$ s$^{-3}$, which is half an order of magnitude larger than mean GH-value (5). This mean turbulence dissipation rate value is about one order of magnitude larger than open-ocean values from observations well away from boundaries (e.g., Gregg, 1989; Yasuda et al., 2021). In this part of the Mediterranean tides are so small that main internal-wave sources are reduced by 60% compared with the open ocean.

**4.1 A six-day movie of varying stratification**

The central six-day, single-line data overview of SW in Fig. 1 demonstrates vertically variable temperature differences over $\pm0.003$°C with time (Fig. 2). The quasi-3D movie reveals that stratification is advected by waterflow from the side (e.g., day 439) or relatively slowly and wavy-like varying from above (e.g., day 440). The mean flow is directed westward at 0.04 m s$^{-1}$, so that the mooring-array is passed in half an hour, 0.02 day, or in 1 s of movie-time.

The movie starts with almost uniform relatively cool waters. Soon, slightly warmer waters are seen entering sideways near the top from the northeast. The faint flickering of light-blue colours represents turbulent motions down over several T-sensors and over several lines, but this is limited in appearance due to the relatively large overall temperature range. On day 439 fast rugged turbulent clouds enter from the east, under warmer waters above. Day 440 shows dominant waves agitating the warmer waters above, with strong variations between fast and slow waves, in trains and solitary passages. Roughly one inertial period has passed since the previous warming. Towards the end of the day short-term flickering reflects standing-wave motions, probably associated with parametric instabilities of interacting propagating primary internal waves. On day 441, fast turbulent lighting is observed below less warm, but still energetic waters above. Such lighting is also observed near the seafloor. Waves not only move from east to west, but occasionally also to the north. In contrast, on day 442 more quiet waves pass. The movie ends with some light clouds barely visible.

Resuming, advection seems more important than anticipated as warming is not exclusively found from below or above. Of course, precise sources are unknown as advected waters can be heated from above or below elsewhere. The amount of turbulence varies strongly over short periods of 0.02-0.4 day. No clear periodicity is observed, although some indication is found for variability of approximately inertial timescale. The deep sea is seen never to be stagnant.

For comparison, 0.5-day, one-thirds slower movies of data details are given in Section 4.2 and in Appendix A.

**4.2 Short-period details from a single line**

During 2.5 days of rather steady waterflow with speed of about 0.04 m s$^{-1}$ (Fig. 1b) in east-southeasterly direction, observed temperature is generally stably stratified, with largest N-values at the beginning of the record followed by weakest values about one mean buoyancy period later (Fig. 3a). Temperature and stratification variations with time range between maximum 2-m small-scale and mean buoyancy periods and have amplitudes of several tens of meters. Around day 437 isothermal contours occasionally move in opposite phase to those above (or below), which evidences local vertical mode-2 variability that is typical for, a.o., parametric instabilities. These instabilities result from standing-wave motins following weakly nonlinear, resonant interactions between freely propagating parent waves that have typically 4-5 times larger scales. Standing waves do not propagate, and extract energy from parent waves up to the point of their breaking (Davis and Acrivos, 1967; Thorpe, 2005). In the present example, they occur at various scales which suggests different sources. The largest have a duration of 0.1-0.2 day, suggesting near-inertial motions as parental wave source. Rugged isothermal contours best represent even shorter periodicities that likely associate with turbulent overturns.

The vertically averaged turbulence dissipation rate also varies over short <0.1 day timescales (Fig. 3b), around its 2.5-day and 124-m average that equals to $1.8\pm0.9\times10^{-10}$ m$^2$ s$^{-3}$. The vertical stratification variation is reflected in differently sloping, but otherwise featureless temperature-variance spectra (Fig. 3c). Most exceptional is the spectrum for the lowest T-sensor, which slopes over two-and-a-half orders of magnitude like $p = -1$ that can be modeled as intermittent signal (Schuster, 1984). This is likely due to alternating GH from below and, GH-blocking, SW from above.

In the 31-records smoothed mean for the lower half of T-sensors, half an order of magnitude less variance is found than for the upper half, with largest difference between the two spectra around 20 cpd and smallest difference at <2 and >1000 cpd. Between frequencies $N_{max} = 2.4N_m \leq \omega \leq 1000$ cpd for the lower half, the spectrum significantly adopts BO-scaling along (unscaled) spectral slope p = -7/5. Thus, apart from the lower h = 2 m above seafloor, motions seem dominated by convection turbulence over about half of the 124-m vertical range, and between small-scale internal wave and white noise frequencies. At lower frequencies, a steep (unscaled) p < -2 slope reaches to $2\Omega$, and with a suggestion of BO-scaling in the IGW band.

For the upper half between $N_{max} = 2.4N_m \leq \omega \leq 100$ cpd, IW-scaling along (unscaled) p = -2 is observed. As this range is outside the IGW-range, it either suggests dominance of (unresolved) < 2-m small-scale internal waves, or, more likely, finestructure contamination of rapid advection of finite temperature variations passed the T-sensors, possibly due to standing-wave instabilities. This puzzling slope-dependence may also reflect incomplete, overturns lager than the 60 m vertical averaging range. Arguably, the IW-slope can be extended across 4-5 cpd $\approx N_m \leq \omega < 2.4N_m$ including a BO-scaled dip around $2N_m$, before the spectrum suggests resuming BO-scaling at IGW frequencies. At $100 < \omega < 500$ cpd, the spectrum aligns with KO-scaling along (unscaled) p = -5/3 reflecting inertial subrange before bending to BO-scaling. The average between the lower- and upper-half spectra, i.e. the mean over 124 m, shows KO-scaling for $2N_m \leq \omega \leq 500$ cpd, with a small depression around 100 cpd. Overall, shear-turbulence seems dominant in stratification overturning.

During a one-day period of dominant stratification in the upper half (Fig. 4), several differences are observed with respect to Fig. 3. The waterflow measured at h = 126 m above seafloor is steady at a speed of 0.05 m s$^{-1}$, but in westerly direction. An about 0.1-day periodicity is observed in relatively warm waters above (Fig. 4a). Small-scale GH is seen near the seafloor between days 442.8 and 443.15. The vertically averaged turbulence dissipation rate (Fig. 4b) is largest during this GH-episode. The one-day mean value is $1.6\pm0.8\times10^{-10}$ m$^2$ s$^{-3}$, with about one-third above mean GH-value (5) attributable to breaking IGW, most likely via parametric instability in the upper half. Although the shortness of record does not resolve the IGW band, one-day mean spectra (Fig. 4c) are vertically more uniform than in Fig. 3. Between about $2N_m < \omega < 150$ cpd spectra from all heights slope along BO-scaling, with largest temperature variance at the T-sensors in the upper half of the mooring-array as before. While in the lower half BO-scaling continues outside this range to lower and higher frequencies up to 1000 cpd, that of the lowest T-sensor adopts the Im-scaling while the upper-half spectrum aligns with KO-scaling of inertial subrange, up to 1000 cpd before deflecting to noise. At lower frequencies a short steep increase in variance is observed in the upper-half spectrum. While the extended BO-scaling of convection turbulence for the lower half may reflect dominant GH, its partial occurrence in the upper half may either reflect a response to GH from below or some interaction between GH and parametric instabilities, since both do not propagate like free waves.

Associated with Fig. 4a, a half-day real-time, 72-s movie demonstrates the GH-dominant middle portion of the one-day episode. This movie is three times slower than that of Fig. 2 so that waterflow at a speed of 0.05 m s$^{-1}$ passes the mooring-array in about 2 s. It shows lower-half details by more focused colouring and masking the upper-half stratification by shrinking the colour range. The movie indicates relatively slow but active short-term motions, initially foremost originating from the seafloor up but also in the stratified layer above. Different motions are visible, with prominent quasi-interfacial waves at the border to masked stratification above, alternated with vertically up- and down-going quasi-standing motions varying between neighbouring lines, both at the interface and near the seafloor. Towards the end of the movie, stratification is moved up and larger puffs of GH-warmed waters move up, besides some limited horizontal advection of clouds of warmer water heated or overturned elsewhere.

A half-day example of large turbulence values and a sudden passage of warm column at all levels is shown in Fig. 5. The waterflow is steady directed westward at a speed of 0.015 m s$^{-1}$, three times less than for Figs 3, 4. The overall mean turbulence dissipation rate equals $1\pm0.5\times10^{-9}$ m$^2$ s$^{-3}$, <10% difference between Ellison (1957) and Thorpe (1977) method values, well to within error as for Figs 3, 4. Concentrations of several large turbulent overturns are found around days 441.35 and 441.55. This mean turbulence dissipation rate is half to one order of magnitude larger than in Figs 3 and 4. Accordingly, temperature variance is larger as well, at all heights. Like in Fig. 4, the lower-half spectrum adopts BO-scaling across its entire frequency range up to the 3000-cpd lpf cut-off. The lowest T-sensor data are anomalous, with BO-scaling between about $50 < \omega < 1000$ cpd and a hump of variance-increase around 20 cpd $\approx N_{max}$. Also in the upper-half spectrum BO-scaling is evident, at $\omega < 100$ cpd, with KO-scaling found between 120 < ω < 1000 cpd, and probably extended to 3000 cpd after application of the vertical lpf as is done for the 3-123 m spectrum. While this short period shows most extensive occurrence of BO-scaling, the associated convection turbulence is not mainly induced by GH. No evidence is found for GH in the temperature image and average turbulence dissipation rate is nearly ten times larger than that of mean GH. Instead, the turbulent overturning is generated elsewhere, partially from above, and advected passed the mooring-array. Question is whether it is related with larger-scale sub-inertial motions.

Like Fig 4a, a movie can be activated of the length of Fig. 5a. The movies have the same set-up, but the one in Fig. 5a is under three-times slower waterflow of 0.015 m s$^{-1}$, measured at h = 126 m. Despite the slower upper-layer waterflow, the movie of Fig. 5a shows much more activity than that of Fig. 4a, with turbulent clouds rapidly passing rather than up- and down-going motions, especially in the middle of the movie. The passing of turbulent clouds lasts less than 1 s, i.e. < 600 s in real-time, at a speed that is nearly one order of magnitude larger than the particle speed (measured at h = 126 m). The turbulence clouds represent tens of meters tall overturns. Standing motions are barely observed, except for some towards the end of the movie.

**4.3 Increased spectral smoothing**

While the temperature variance spectra of Figs 3c-5c show reasonably uniform and consistent slopes, they were calculated in smoothed form for a single-line T-sensor mooring, without taking advantage of the multiple lines of the large-ring mooring. Further improved statistics for records from the 17-day SW period is investigated in this sub-section.

Applying spectral smoothing over vertically three T-sensors, if independent, and all 45 lines for two 7-day periods demonstrates the variability in random statistics per frequency range (Fig. 6). The spectra, now presented in unscaled form for non-filtered data, show a twice wider graph-spread (on logarithmic scale) between about 0.6 < ω < 60 cpd compared to the rest of the spectrum. This suggests that the internal-wave and large-scale turbulence ranges are not entirely resolved by the array's horizontal scales O(10) m, i.e. motions at these scales do not obey improved statistics for quasi-random signals. Thus, the ~120 independent data records per spectrum sample partially dependent, the same, signals. Nevertheless, the spectra support some findings of temperature spectra Figs 3c-5c. The temperature-variance spectra are compared with the mean spectra from three current-meter records measured at h = 126 m above seafloor, for kinetic energy KE and horizontal waterflow difference dU. While waterflow spectra are generally too noisy for $\omega > 2\Omega$, temperature-variance spectra are only considered for $\omega < 1000$ cpd to avoid effects of short-term bias and noise, as resulted from evaluation of vertically mean and upper-half spectra in Figs 3c-5c.

The two one-week periods distinguish increased upper-layer variance in temperature (h = 109 m) during the first week. The increased variance extends to sub-inertial frequencies starting at $\omega \approx \omega_{min} \approx 1$ cpd. No peak is observed at f, also not in KE, which demonstrates elevated levels at sub-buoyancy frequencies only. This suggests that the driving motions are more likely residing at sub-mesoscales and not at IGW, unless local relative vorticity is substantial to affect $f_{eff}$. During the second week, temperature variance is reduced near f and elevated around N, at all heights. No other clear peaks and no gaps are observed, which potentially allows investigation in terms of spectral slopes across finitely-wide frequency bands.

In the limited waterflow spectra, no obvious consistent slope-scaling is apparent. A suggestion is found in both KE-spectra for Im-scaling along p = -1 indicative of intermittency at sub-inertial frequencies $\omega < f$, and a steep slope p < -2.5 across the IGW. The difference and correspondence between the two one-week spectra is extended to super-IGW frequencies for temperature variance.

In Fig. 6a, the wider band between $10 < \omega < 100$ cpd of upper T-spectrum distinguishes from those of lower- and mid-heights. The upper spectrum aligns with KO-slope p = -5/3, not accounting for the narrowing in variance-variation around $2\Omega$ and 100 cpd, also up to 1000 cpd. Between $10 < \omega < 100$ cpd at lowermost T-sensors, the spectrum aligns with Im-slope p = -1, while at $100 < \omega < 1000$ cpd and for mid-heights at $10 < \omega < 1000$ cpd it aligns with BO-slope p = -7/5. With the notion that the vertical resolution is limited to 2 m, the band $10 < \omega < 100$ cpd appears like a transition between turbulent overturns and small-scale internal waves as it includes the maximum 2-m small-scale internal wave frequency. For $\omega_{min} < \omega < 2N_m$, the lower- and mid-height spectra roughly align with IW-slope p = -2.

In Fig. 6b, a week with half the stratification and 20% smaller turbulence dissipation rate compared to that of Fig. 6a, see Fig. 1, the aligning with IW-slope extends to 0.3 cpd < $\omega$ < $\omega_{min}$, and includes a small peak near N at all heights. Variance variations are still considerable and "errors" in slope-determination are large. For mid-heights at $\omega$ < N and other heights at $\omega$ < $\omega_{min}$, peaks align with BO-slope. However, the level of temperature variance is an order of magnitude elevated above that of BO-slope alignment at $\omega$ > 100 cpd. This difference in levels transits via the IW-slope, but the relationship between internal waves and convection turbulence is not immediately obvious.

Extended smoothing and vertical coherence are investigated in Fig. 7, under the notion that considerable differences in stratification and temperature-variance distribution are observed as a function of height above the seafloor. In this figure, smoothing is applied over 17 days and approximately 2000 independent T-sensor (pairs). The average N ≈ 2.2f = 2.0$f_h$, and $N_m$ ≈ 4$f_h$.

While temperature is highly coherent over 2-m vertical scales at sub-inertial frequencies, significantly incoherent signals are found at approximately $\omega$ > 1000 cpd (Fig. 7a). The incoherence range is also visible in phase difference, which is significantly away from zero at $\omega$ > 500 cpd (Fig. 7b), and in variance-width reduction with respect to that of random noise $\omega$ > 300 cpd (Fig. 7c). Coherence is still >0.8 at overall maximum 2-m-scale buoyancy frequency $N_{max}$, which suggests that stratified turbulent motions can have some coherence and/or that freely propagating internal waves may exist supported by unresolved finer vertical-scale Δz < 2 m stratification. It is noted that for $\omega$ > $N_{max}$, the coherence and associated temperature-variance spectra are somewhat affected by short-term bias, as modest vertical lpf could not be applied on these data (the cut-off being on two-times larger scales than the 2-m distance between T-sensors). Under SW-conditions, the effect of short-term bias is noticeable in spectra at about $\omega$ > 500 cpd (Fig. 7c).

As suggested for smaller-period data records in Figs 3c, 5c, 6a, the smoothed temperature-scalar spectrum from the 17-day SW period tends to align with KO-scaling of p = -5/3, at $N_{max}$ < $\omega$ < 500 cpd, and likely beyond < 1000 cpd. Larger variance width extends across IGW to sub-mesoscale sub-inertial frequencies and aligns approximately with BO-scaling of p = -7/5 between about 0.06 cpd < $\omega$ < N ≈ 2$f_h$. Although KO-scaling is reflective of up-frequency energy cascade of predominantly shear turbulence, it is observed across a significantly coherent part of the spectrum. Turbulence is therefore unlikely fully isotropic, or if so shows >2-m large scales. While this is plausible in the weak stratification found in the deep Mediterranean, the coherent portion of turbulence may also reflect anisotropy.

Anisotropy is the basis of 'stratified turbulence', which is also part of convection turbulence following Bolgiano (1959). Although the present data suggest a transit across IGW between anisotropic turbulence and small-scale internal waves and sub-mesoscale motions, longer timeseries are required for further evidence on interaction between these motions and, possibly, on direction of energy transfer.

## 5 Discussion and Conclusions

Future studies using yearlong data records will elaborate on the possible energy transfer between (sub-)mesoscale, IGW, and turbulence motions. Inevitably, such studies will contain a mix of SW conditions, studied here, and near-homogeneous conditions, investigated elsewhere (van Haren, 2026 submitted). The distinctively different conditions occupy about 40% and 60% of yearlong records, respectively. While under near-homogeneous conditions turbulence is mainly governed by GH from below, two additional turbulence sources are recognized under SW conditions.

The parametrically excited instabilities of interfacial internal waves will affect thin stratified layers and shear. They are found abundant, but have limited diapycnal mixing effect as their average turbulence dissipation rate is about $3\pm1\times10^{-11}$ m$^2$ s$^{-3}$, half an order of magnitude smaller than that of GH. This confirms open-ocean turbulence dissipation rate values observed away from boundaries using high-precision shipborne instrumentation (Yasuda et al., 2021). Probably, this internal-wave breaking mechanism explains observations by Ferron et al. (2017). The small but turbulent exchange yields diffusive interfaces. In movies it shows like highly local vertical exchange with limited advective transport, like stratification-suppressed GH.

The second turbulence source under SW conditions has tenfold larger mean turbulence dissipation rate values of about $4\pm2\times10^{-10}$ m$^2$ s$^{-3}$, half an order of magnitude larger than that of GH. This mechanism has unlikely been observed by Ferron et al. (2017). It associates with stratified waters of typically N = $2f_h$. Despite this stratification being well larger than that under near-homogeneous conditions, it reflects marginally-stable conditions given the large turbulence value. With the warmer waters, it most likely associates with a larger scale eddy, initiated at near- or sub-inertial (sub-meso) scale. Convection turbulence becomes tilted away from Earth's rotational vector outside polar regions. At mid-latitudes, convection turbulence, or 'slantwise' convection (Marshall and Schott, 1999; Straneo et al., 2002), along eddy tubes tilted by the horizontal Coriolis parameter may appear as stable stratification in local vertical (z) direction. However, following stability analysis (van Haren, 2008) this stratification will be marginally stable for large-scale linear background shear at buoyancy frequency of $N_{lin} = 2f_h$, and for smaller scale shear by transients associated with nonlinear stability at $N_{nl} = f_h, 4f_h$. The condition of large-scale $N = 2f_h$ has been observed previously in association with small-scale vertical motions that were found coherent over at least 50 m in the deep Mediterranean (van Haren et al., 2014). In the present observations it also associated with large 100-m scale N, while mean 2-m small-scale $N_m = 4f_h$.

The 3D movies from the nearly 0.5-hm$^3$ array do not resolve the tilted eddies, albeit that the warmer waters are seldom being advected uniformly over the vertical: their fronts always appear slightly tilted. At a scale O(100) m the typical eddy size is poorly resolved. Given the observed local particle speed, the horizontal scale of an inertial motion is about 1 km, while that of 10-30 day variable meso-scale motions is 25-75 km under steady 0.03 m s$^{-1}$ waterflow. Puzzling in the 3D-movies are the rapidly passing turbulence clouds, which reflect overturning elsewhere, but of which development is difficult to trace in vertical direction. Apparently at the size of the large-ring mooring, the passing of turbulence clouds is how slantwise-convection turbulence shows. While the limited size of our 3D mooring hampers direct investigation of coupling of turbulent clouds with sub-inertial eddies, their overturning is clearly different from stagnant waters.

Indirectly, a coupling between sub-inertial and turbulent motions is suggested from spectral analysis. With improved statistics in the turbulence range, a distinction is made between inertial and buoyancy subranges, in frequency but also in time and in the vertical. The buoyancy subrange evidences importance of convection turbulence, apparently also in slantwise direction that is maintained under thermal wind (shear) balance.

Technically, both under near-homogeneous conditions and, only slightly more pronounced, under SW conditions, the frequency-dependent effects of spectral smoothing demonstrate the complex statistics of oceanographic data. Under SW, only at $200 < \omega < 500$ cpd spectral smoothing acts like for quasi-random signals, at vertical scales < 2 m and horizontal scales < 9.5 m. While band-smoothing, averaging of neighbouring spectral bands, reduces the vertical power-range, the effect of poorly resolved statistics extends beyond $\omega < \omega_{min}$ and $\omega > N_{max}$. The variance-width is largest near mean maximum small-scale buoyancy frequency $N_m$ and seems to reduce again at sub-inertial frequencies. This relates non-random statistics to largest turbulent overturns, but especially also to small internal-wave scales.

*Data availability.* Only raw data are stored from the T-sensor mooring-array. Analyses proceed via extensive post-processing, including manual checks, which are adapted to the specific analysis task. Because of the complex processing the raw data are not made publicly accessible. Current meter and CTD data are available from van Haren (2025): "Large-ring mooring current meter and CTD data", Mendeley Data, V1, https://doi.org/10.17632/f8kfwcvtdn.1. Movies to Figs 2, A1 and A2 can be found in van Haren, Hans (2026), "Movies to: Deep Mediterranean turbulence motions under stratified-water conditions", Mendeley Data, V1, https://doi.org/10.17632/b8jv8378dj.1. Atmospheric data are retrieved from https://content.meteoblue.com/en/business-solutions/weather-apis/dataset-api.

*Competing interests.* The author has no competing interests.

*Acknowledgments.* This research was supported in part by NWO, the Netherlands organization for the advancement of science. Captains and crews of R/V Pelagia are thanked for the very pleasant cooperation. The team of ROV Holland I performed an excellent underwater mission to recover the instrumentation of the large ring. NIOZ colleagues notably from the NMF department are especially thanked for their indispensable contributions during the long preparatory and construction phases to make this unique sea-operation successful. The KM3NeT research infrastructure is being built, operated and maintained by the KM3NeT Collaboration, and funded by a large number of local, national and international agencies. I am indebted to colleagues in the KM3NeT Collaboration, who demonstrated the feasibility of deployment of large three-dimensional deep-sea research infrastructures.

**Appendix A Movies too**

The choice of colour palette and range are crucial for visualization of imaging, and also for movies. The long movie of Fig. 2 appears somewhat boring due to the masking of turbulent motions in limited colour ranges. The shorter movies of Figs 4 and 5 addressed some details of weakly turbulent GH and more intense, slantwise-advected turbulent overturning, respectively. Below, two more detailed periods are addressed, mainly focusing on weak internal-wave turbulence from above including parametric instabilities.

Both half-day SW periods are characterized by vertical temperature difference of about 0.0025°C, relatively large mean waterflow speed of 0.05 m s$^{-1}$, albeit in opposite directions, and turbulence dissipation rates between $2\times10^{-11}$ and $2\times10^{-10}$ m$^2$ s$^{-3}$. The movies show somewhat slow motions, especially when compared with that of Fig. 5a.

The upper-layer mean flow of the SW period in Fig. A1 is directed westward like in Figs 4a and 5a. In the first half, upper-layer warm waters enter from the south and turbulent overturning is largest with dissipation rates averaging about $2\times10^{-10}$ m$^2$ s$^{-3}$. Some internal-wave generated turbulent overturning from above have scales of several tens of meters. This convection-like turbulence is advected out of the array by the mean flow. Turbulent activity diminishes in the second half, and local internal-wave induced up- and down-going breaking seems mainly due to parametric instability, while turbulence dissipation rate decreases to about $4\times10^{-11}$ m$^2$ s$^{-3}$.

Local vertically up- and down-moving turbulence prevails in Fig. A2, as if horizontal advection is negligible, despite the continuous speed of 0.05 m s$^{-1}$ flow, which is directed eastward. Motions are smooth, but variable in time. The up-down quasi-standing local mode-2 motions are again presumably induced by parametric instability. Colour bands are seen to stretch and shrink per line, with the occasional colour-band breaking reflecting local turbulent overturning. Their turbulence is of smaller scales < 20 m and not advected by the mean (upper-layer) waterflow as in the first half of Fig. A1. The average turbulence dissipation rate amounts about $3\times10^{-11}$ m$^2$ s$^{-3}$, commensurate excess turbulence dissipation rate above mean GH-value (5) in Fig. 4.

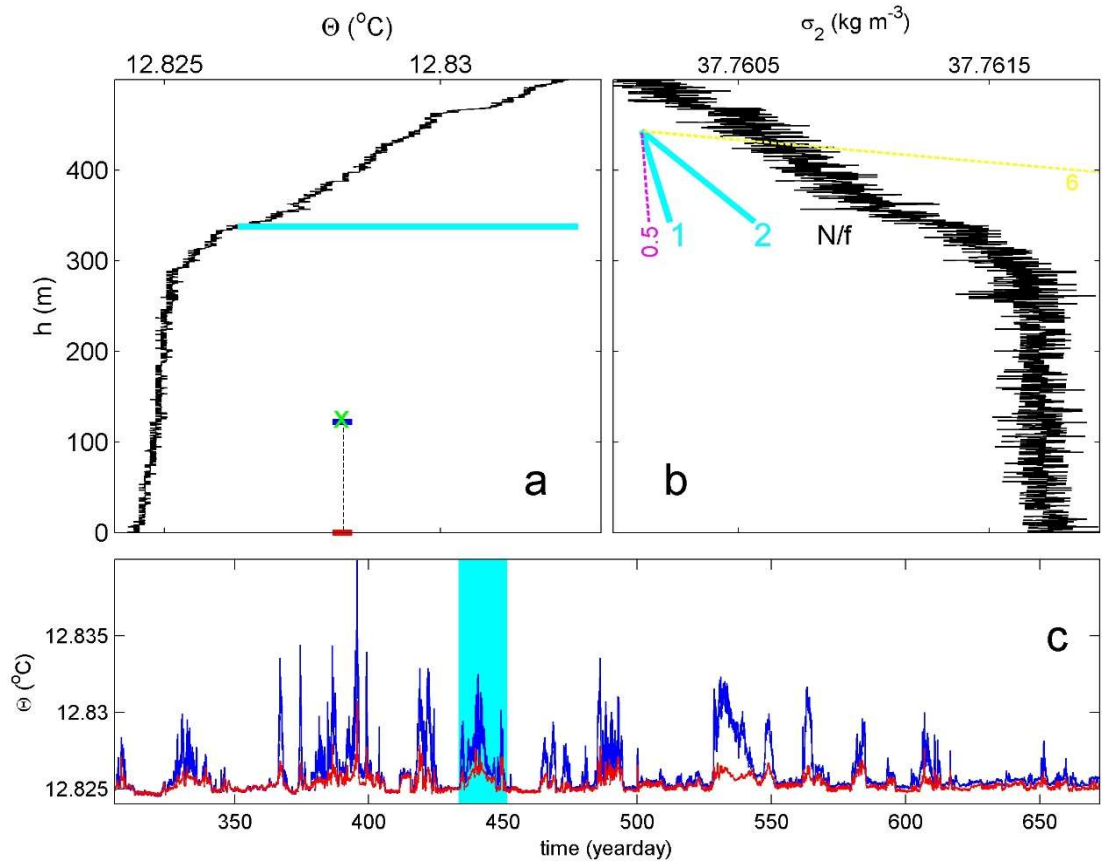

**Figure 0.** Conventional oceanographic data overview from single shipborne Conductivity Temperature Depth 'CTD' profile and from single mooring line in the area of the large-ring. (a) Conservative Temperature (IOC et al., 2010). Lower 500 m of CTD-profile, with indications for height above seafloor of moored instruments: current meter (x), uppermost (blue) and lowest (red) moored temperature sensor. The cyan bar indicates the temperature range of the moored instrumentation for the 17-day period highlighted in c. (b) Corresponding density anomaly referenced to a pressure level of $2\times10^7$ N m$^{-2}$. Several 100-m scale vertical density-stratification rates are indicated in terms of the ratio of buoyancy over inertial frequency N/f. The cyan lines indicate the ratios corresponding to those of the highlight in c. (c) Yearlong time series of detrended Conservative Temperature from uppermost and lowest T-sensors of mooring line 24. Time is given in days of 2020 (+365 in 2021). The box highlights temperature time series under stratified-water 'SW' conditions discussed in this paper.

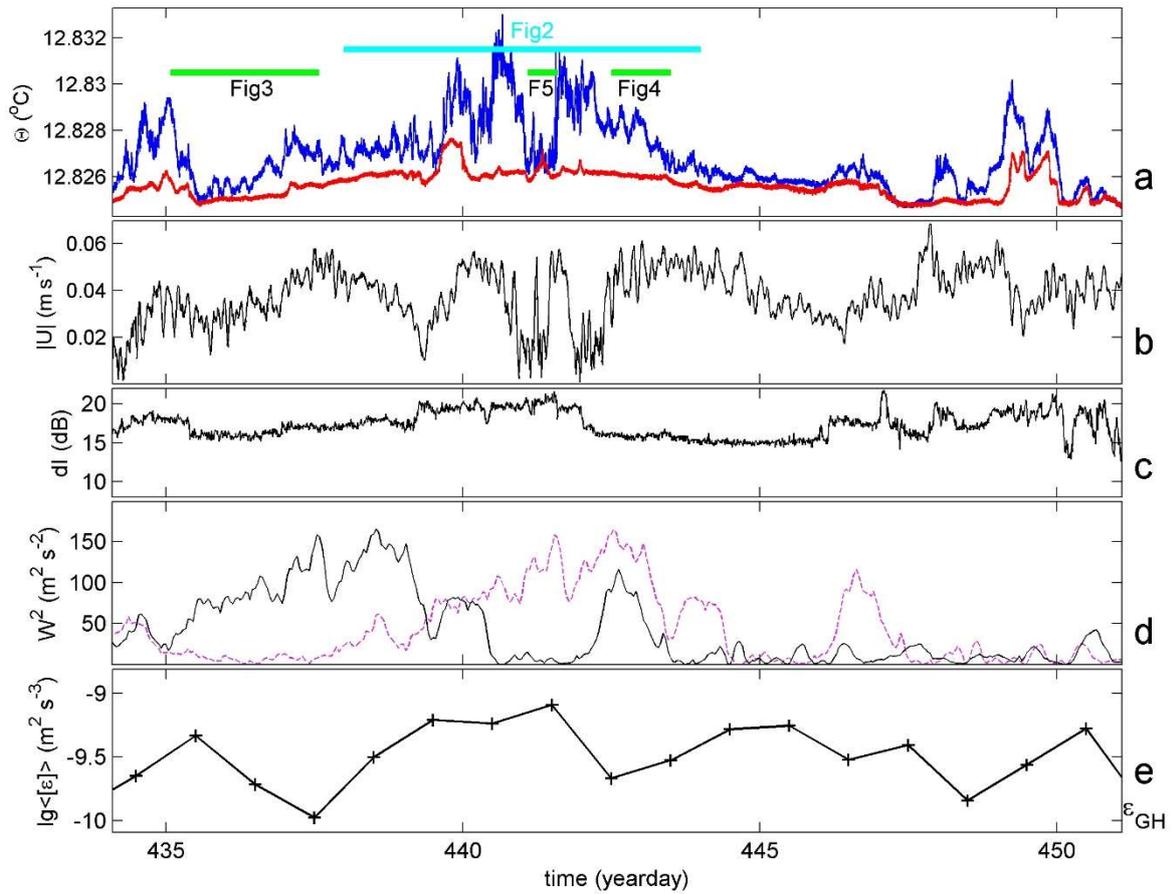

**Figure 1.** Time-series overview of several environmental parameters during a 17-day period under SW conditions at the large-ring mooring. (a) Conservative Temperature from line 24, T-sensors at h = 0.7 (red) and 124.7 m (blue) above seafloor. Unfiltered, bias-corrected via de-trending, and 10-s sub-sampled data. Horizontal colour bars indicate time periods of Figs 2-5. (b) Waterflow amplitude measured at h = 126 m, hourly low-pass filtered 'lpf'. (c) As b., but for relative acoustic-echo amplitude. (d) Wind variance measured at island-station 'Porquerolles', about 20 km north of the mooring. In magenta, four days shifted forward. (e) Vertically and daily averaged turbulence dissipation rate using Thorpe (1977) method. The level of mean geothermal heat 'GH' dissipation rate is indicated.

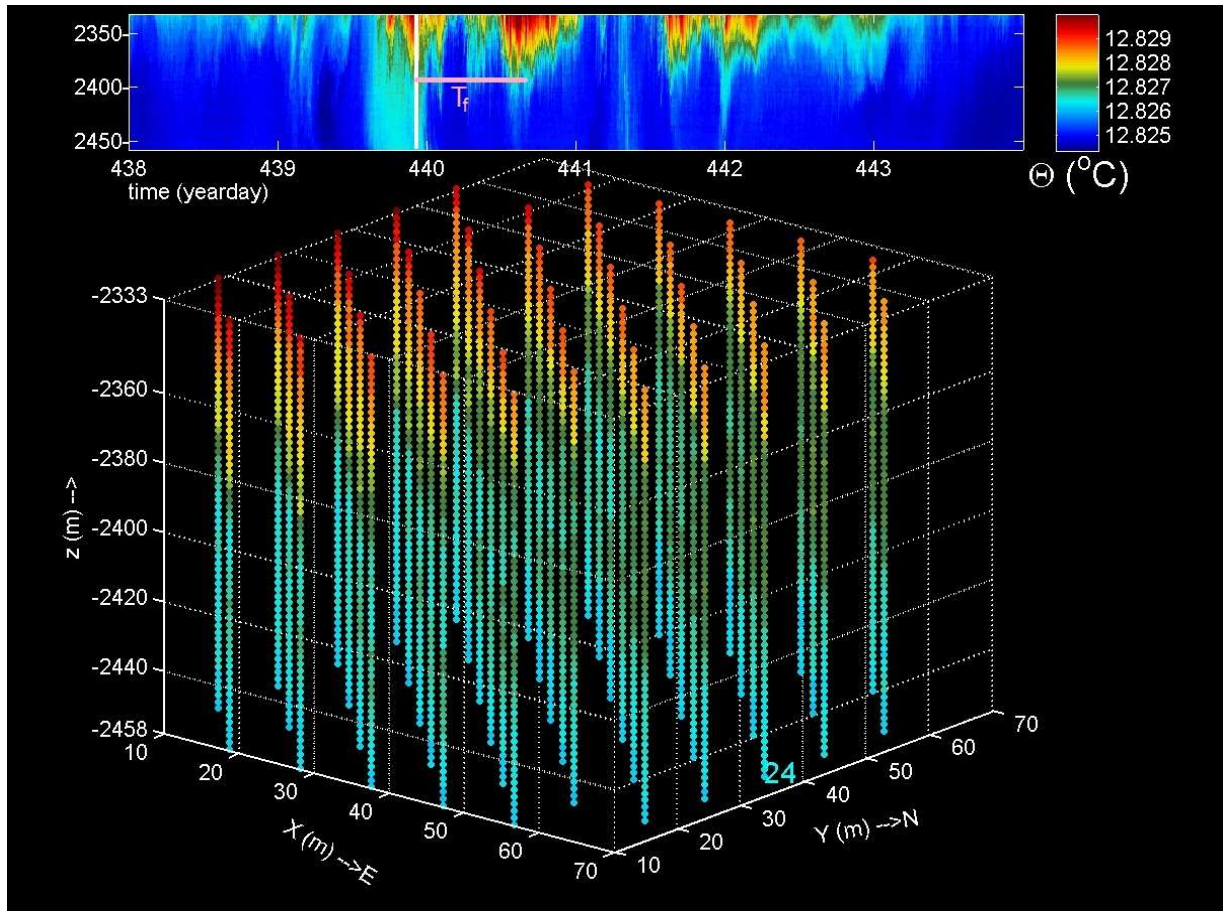

**Figure 2.** Quasi-3D movie from six days of 3000-cpd (cycles per day) lpf temperature data from about 2800 T-sensors in nearly 0.5-hm$^3$ mooring-array. Each sensor is represented by a small filled circle, of which the colour represents Conservative Temperature in the scale above, which covers a total range difference of 0.005°C. In the movie, above the cube, which is vertically depressed by a factor of two relative to horizontal scales, a white time-line progresses in a 6-d/124-m time/depth image from line 24 on the east-side of the cube. In the image above, the horizontal magenta line indicates one inertial period. The 288-s movie is accelerated by a factor of 1800 with respect to real-time.

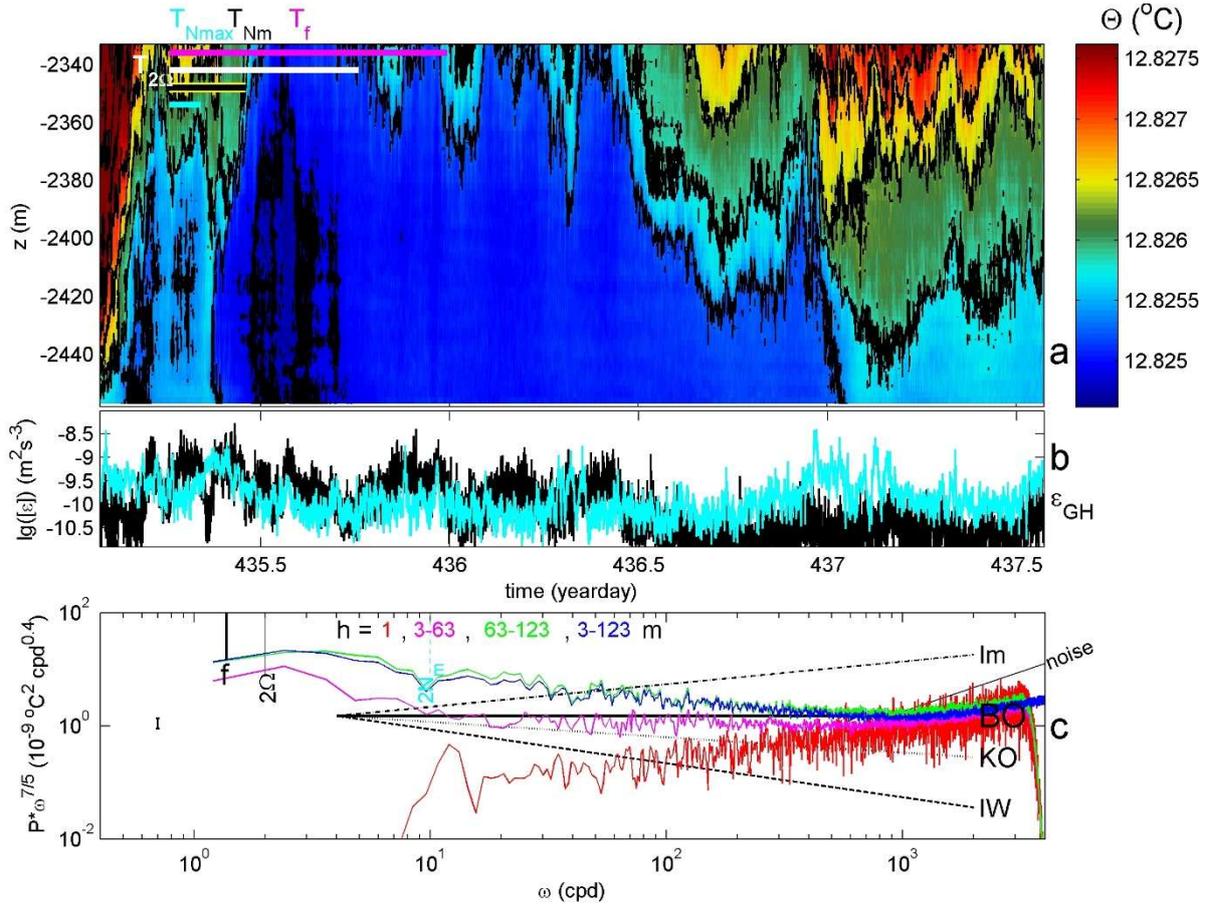

**Figure 3.** Detail of 2.5-day moored T-sensor data from corner-line 47 during a moderately turbulent SW-period of Fig. 1. (a) Time/depth series of 3000-cpd and 0.2-cpm lpf Conservative Temperature. Horizontal bars indicate one inertial period $T_f$, one semidiurnal period $T_{2\Omega}$, one mean of maximum 2-m small-scale buoyancy periods $T_{Nm}$, and one maximum small-scale buoyancy period $T_{Nmax}$. The mean 124-m large-scale buoyancy period equals $T_N \approx 1.5T_f, \approx 2\Omega$. Ellipses indicate several parametric instabilities around day 437. Over the entire colour range, the temperature difference amounts $\Delta\Theta = 0.003°C$. Black contours are drawn every 0.0005°C. (b) Time series of 124-m vertically averaged turbulence dissipation rate calculated using Ellison (1957) method (cyan) with cut-off frequencies at $2N_m \approx 10$ cpd and 3000 cpd. The [ε] calculated from Thorpe (1977) method is given in black. (c) Weakly smoothed (about 10 degrees of freedom 'dof') spectrum of [10, 3000] cpd band-pass filtered data of lowest T-sensor (cyan) in comparison with 3000-cpd lpf heavily smoothed (300 dof) lower (magenta) and upper (green) layer spectra, and with 900 dof (cf. error bar) 120-m averaged 0.2-cpm lpf spectra (blue). All spectra are scaled with BO-slope p = -7/5 for active-scalar quantities. Several spectral slopes are given by straight lines, and indicated by their abbreviations (see text). Given are inertial frequency f, semidiurnal $2\Omega$, and $2N_m$.

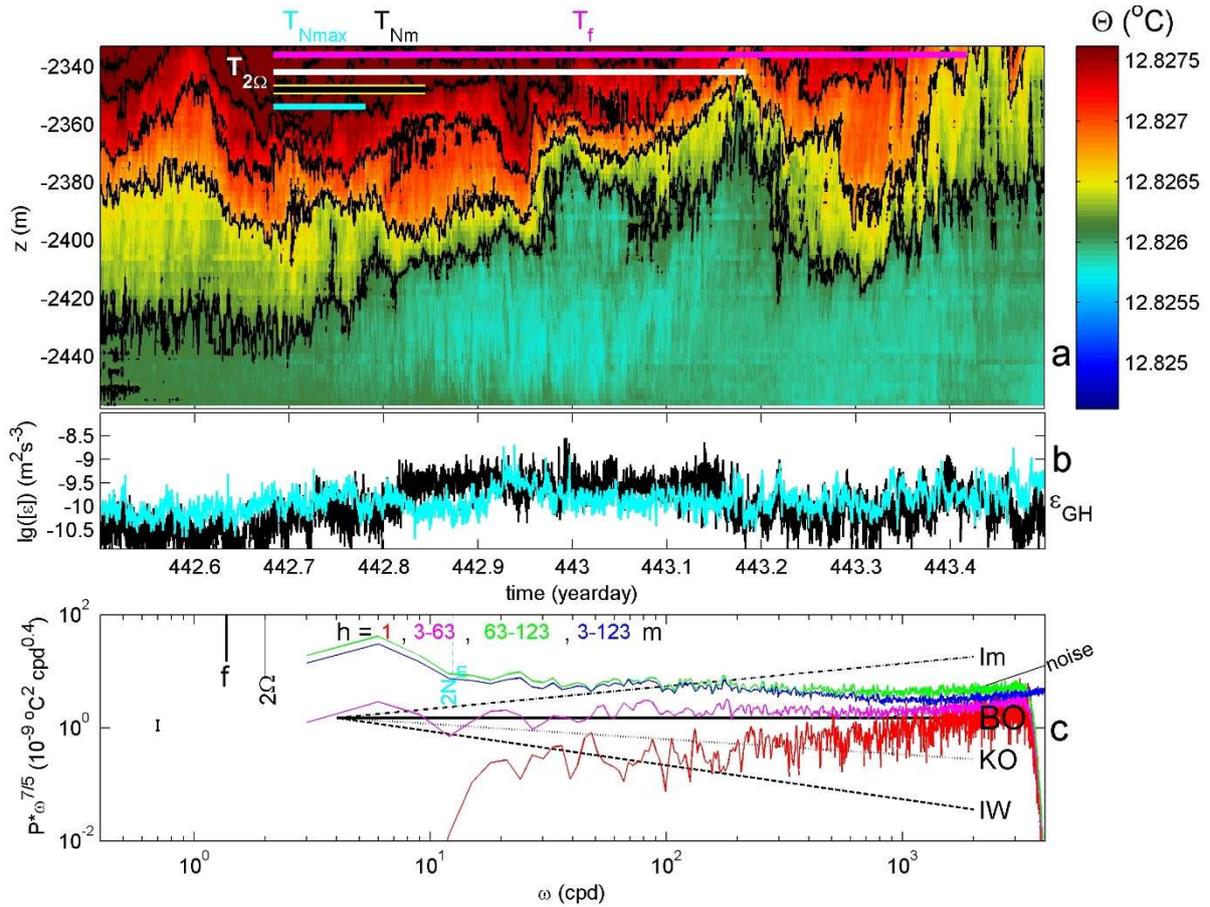

**Figure 4.** As Fig. 3, but for 1.0-day example of SW with some suppressed GH activity near the seafloor. In a., black contours are drawn every 0.0005°C. A 72-s movie can be activated which covers the half-day of data between days 442.8 and 443.3. In the movie's upper panel contours are drawn every 0.0003°C.

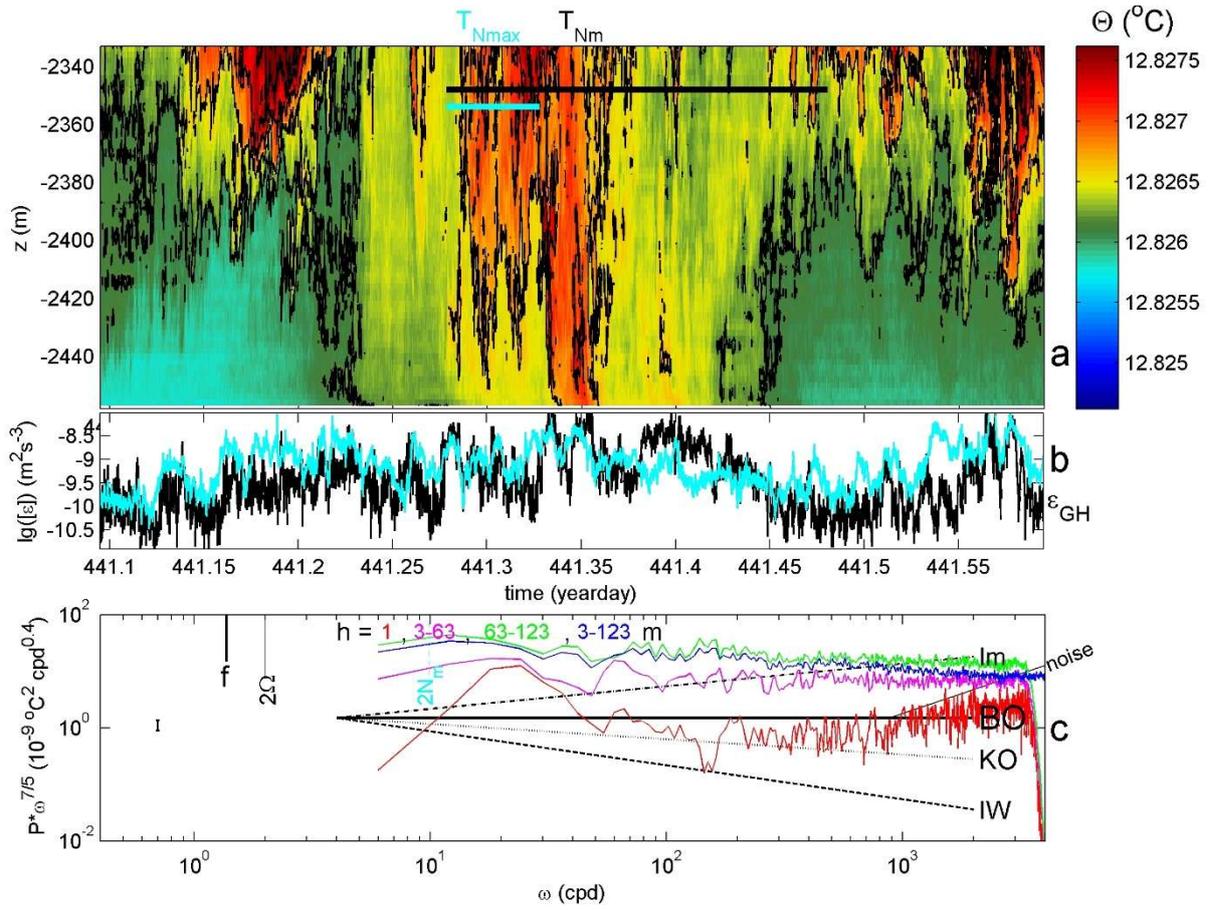

**Figure 5.** As Fig. 4, but for 0.5-day example with strong turbulence and quite variable stratification. A 72-s movie is associated.

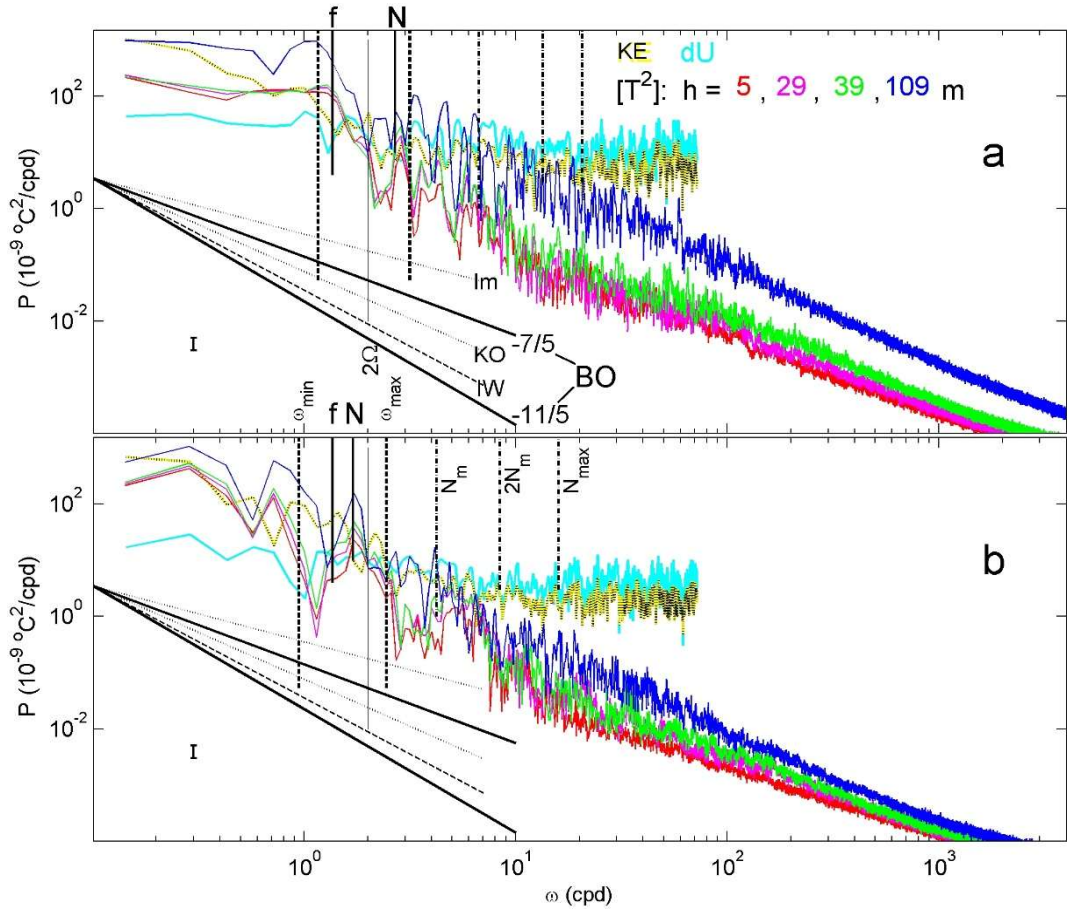

**Figure 6.** Unscaled spectra for two non-overlapping one-week SW periods of Fig. 1. Kinetic energy 'KE' and horizontal waterflow differences 'dU' averaged over the three current meters at h = 126 m (arbitrary vertical scale) are compared with non-filtered temperature variance 'T$^2$' averaged over all 45 lines and three vertically-neighbouring T-sensors, if not-interpolated, around heights above seafloor as indicated. For reference, model spectral slopes and some frequencies are given, see text. (a) Days 436.5-443.5, for which mean N = 2.5f$_h$. (b) Days 444.1-451.1, for which mean N = 1.5f$_h$.

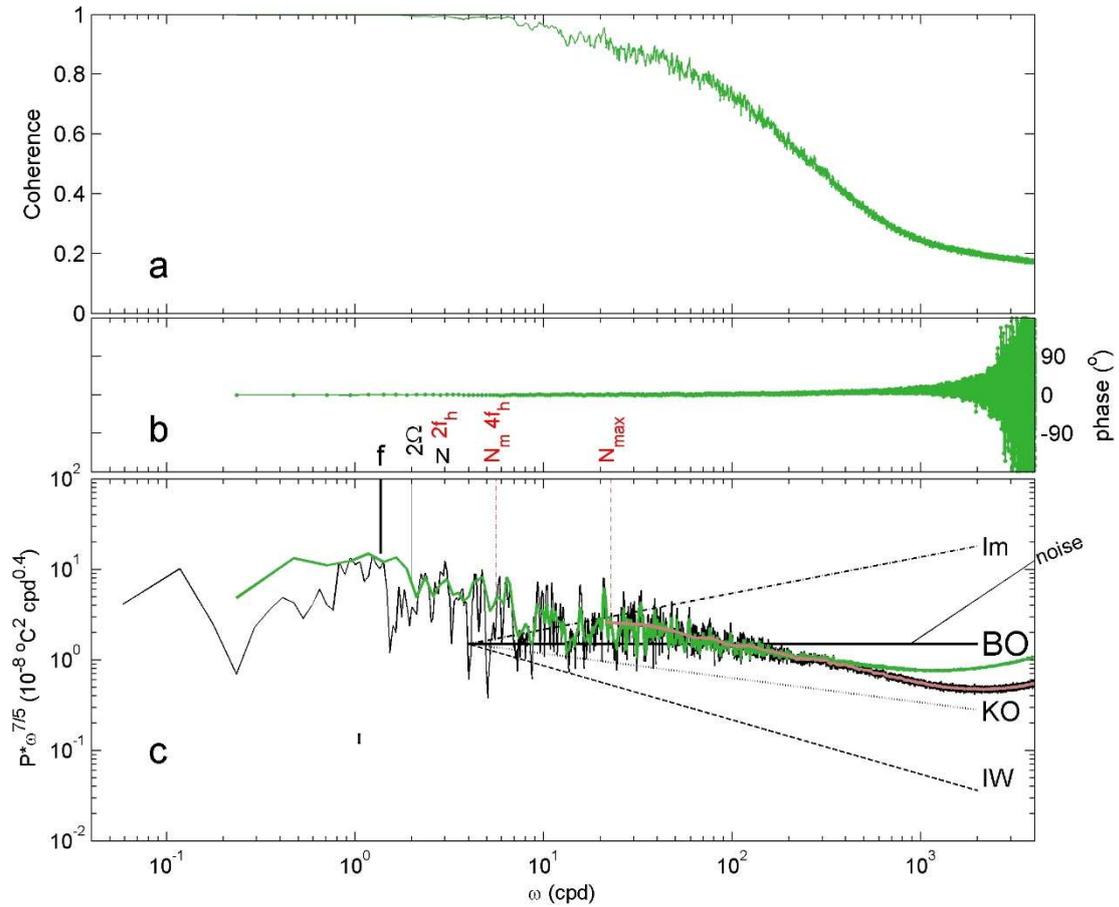

**Figure 7.** Spectral statistics for 17-day period of Fig. 1 for all independent, not-interpolated moored T-sensor data records, excluding from uppermost and lowest T-sensors. (a) Very heavily smoothed (~20,000 dof) coherence over 2-m vertical distances of 1957 T-sensor pairs without vertical lpf application. (b) Corresponding phase. (c) Corresponding temperature-variance, in comparison with average over 2088 weakly smoothed spectra for independent T-sensor data for which the vertical lpf is applied, in full range (black) and band-smoothed (pink). Spectra are scaled with scalar-BO $\omega^{-7/5}$ as in Figs 3c-5c. The mean '$N_m$' and maximum '$N_{max}$' 2-m small-scale buoyancy frequencies are indicated.

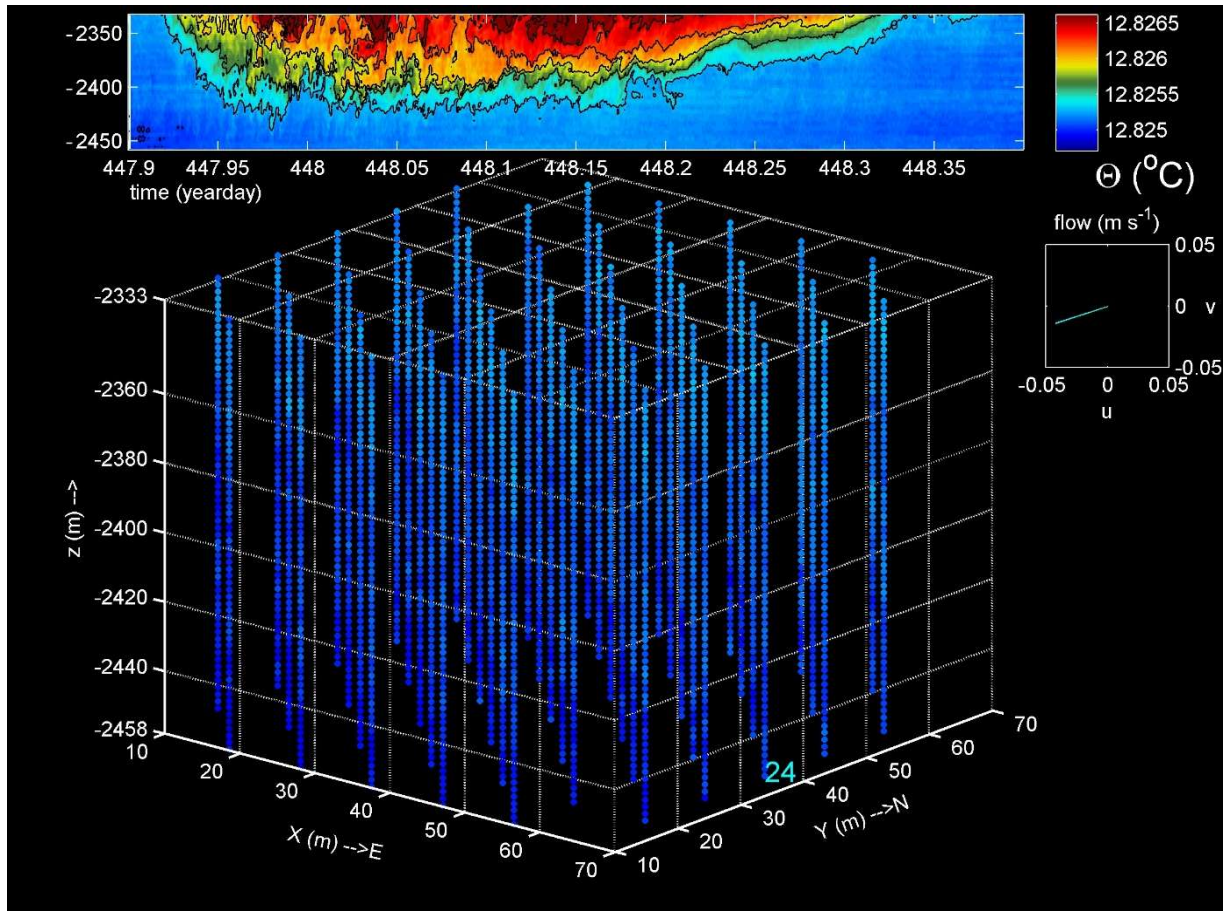

**Figure A1.** As Fig. 2 including movie, but for 0.5-day SW period with internal-wave turbulence from above. In the upper panel black contours are drawn every 0.0003°C. The small panel to the right indicates mean waterflow at h = 126 m.

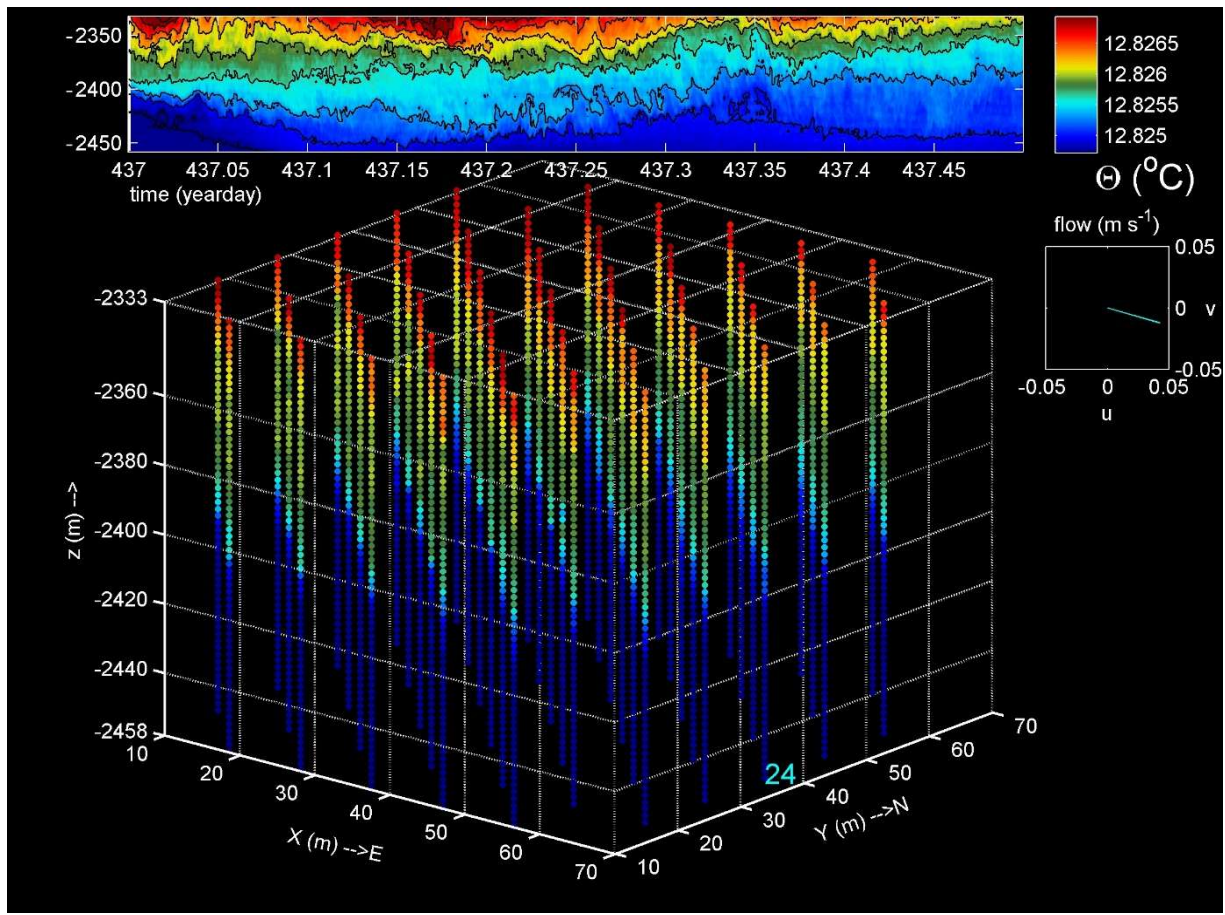

**Figure A2.** As Fig. A1, but for 0.5-day SW period with internal-wave turbulence mainly by parametric instabilities.